\documentclass[12pt]{article}

\usepackage{scicite}
\usepackage{times}
\usepackage{amsmath}
\usepackage{amssymb}

\usepackage{graphicx}    

\newenvironment{sciabstract}{
\begin{quote} \bf}
{\end{quote}}

\title{Ultrafast time-resolved x-ray scattering reveals diffusive charge order dynamics in La$_{2-x}$Ba$_x$CuO$_4$}

\author
{Matteo Mitrano,$^{1,2,\ast}$ Sangjun Lee,$^{1,2}$ Ali A. Husain,$^{1,2}$ Luca Delacretaz,$^{3}$ \\
Minhui Zhu,$^{1}$ Gilberto de la Pe\~{n}a Munoz,$^{4}$ Stella Sun,$^{1,2}$ Young Il Joe,$^{5}$ \\
Alexander H. Reid,$^{4}$ Scott F. Wandel,$^{4}$ Giacomo Coslovich,$^{4}$ William Schlotter,$^{4}$ \\
Tim van Driel,$^{4}$ John Schneeloch,$^{6}$ Genda D. Gu,$^{6}$ Sean Hartnoll,$^{3}$ \\
Nigel Goldenfeld,$^{1,\ast}$ Peter Abbamonte$^{1,2,\ast}$
\\
\\
\normalsize{$^{1}$Department of Physics, }
\normalsize{$^{2}$Seitz Materials Research Laboratory,}\\
\normalsize{University of Illinois at Urbana-Champaign, Urbana, IL 61801, USA,}\\
\normalsize{$^{3}$Department of Physics, Stanford University, Stanford, CA 94305-4060, USA,}\\
\normalsize{$^{4}$SLAC National Accelerator Laboratory, 2575 Sand Hill Road, Menlo Park, CA 94025, USA,}\\
\normalsize{$^{5}$National Institute of Standards and Technology, Boulder, CO 80305, USA,}\\
\normalsize{$^{6}$Condensed Matter Physics and Materials Science Department,}\\
\normalsize{Brookhaven National Laboratory, Upton, NY 11973, USA.}\\
\\
\normalsize{$^\ast$To whom correspondence should be addressed;}\\
\normalsize{E-mail:  mmitrano@illinois.edu, nigel@illinois.edu,}\\
\normalsize{or abbamonte@mrl.illinois.edu.}
}

\date{}

\begin{document}

\baselineskip24pt

\maketitle

\begin{sciabstract}
Charge order is universal among high-T$_c$ cuprates
but its relevance to superconductivity is not established. It is
widely believed that, while static order competes
with superconductivity, dynamic order may be favorable and even
contribute to Cooper pairing. We use time-resolved resonant
soft x-ray scattering to study the collective dynamics of the
charge order in the
prototypical cuprate, La$_{2-x}$Ba$_x$CuO$_4$.
We find that, at energy scales $0.4$ meV $ \lesssim \omega \lesssim 2$ meV,
the excitations are overdamped and propagate via Brownian-like diffusion.
At energy scales below 0.4 meV the charge order exhibits dynamic
critical scaling, displaying universal behavior arising from propagation of topological
defects. Our study implies that charge order is dynamic, so may participate
tangibly in superconductivity.

\end{sciabstract}

One of the key questions in high temperature superconductivity is how it
emerges as hole-like carriers are added to a correlated Mott insulator \cite{Kivelson1998,Vojta2009,Keimer2015}. Soon after the discovery of
Bednorz and M{\" u}ller \cite{Bednorz1986}, it was recognized
that competition between kinetic energy and
Coulomb repulsion could cause valence holes to segregate into periodic structures
originally referred to as ``stripes" \cite{Machida1986,Schulz1986,Zaanen1986}.
Valence band charge order has since been observed in nearly all cuprate
families \cite{Tranquada2004,Abbamonte2005,Kivelson2007,Ghiringhelli2012,Comin2014,daSilvaNeto2014,
LeTacon2011,Chang2012,Berg2009,Fradkin2015,Gerber2015,Hamidian2016},
though it is not known what role, if any, charge order
plays in superconductivity.

It is widely believed that, while static charge order may compete with
superconductivity, fluctuating order could be favorable or even contribute
to the pairing mechanism \cite{Kivelson1998,Kivelson2003,Vojta2009}. It is therefore crucial
to determine whether the charge order in cuprates is fluctuating and, if so, what
kind of dynamics it exhibits.

The generally accepted way to detect fluctuating charge order is to use energy- and momentum-
resolved scattering techniques, such as inelastic x-ray or electron scattering,
to measure the dynamic structure factor, $S(q,\omega)$ \cite{Kivelson2003,Vojta2009}.
This quantity is related to the charge susceptibility,
$\chi''(q,\omega)$, by the fluctuation-dissipation theorem, which asserts
a quantitative relationship between the weakly nonequilibrium dynamics
of a system and its equilibrium fluctuations at finite temperature
\cite{PinesNozieres,LandauLifshitz,Goldenfeld,ChaikinLubensky}.
The time scale of the fluctuations can therefore be inferred from the energy dependence of
the scattering data. The energy scale of charge fluctuations could, however, be of the same order
as the superconducting gap, requiring instruments with sub-meV energy resolution to
detect it. Such x-ray and electron spectrometers do not yet exist,
calling for a different approach.

A way to achieve sub-meV energy resolution is to study the collective excitations
in the time domain.
The effective energy resolution of a time-resolved experiment can be defined as $\Delta
\omega = 2\pi\hbar/T$, where $T$ is the time interval measured \cite{Abbamonte2008}.
Arbitrarily low energy scales can therefore be accessed by scanning to
long delay times. Further, the fluctuation-dissipation theorem guarantees
that the time-domain dynamics of a system may be used to shed light on
its low-energy fluctuations in equilibrium.

When an ordered phase is excited out of equilibrium, its order
parameter could exhibit any of several distinct types of dynamics
\cite{ChaikinLubensky,Goldenfeld,LandauLifshitz}. For example, it might
exhibit inertial dynamics, undergoing coherent oscillation around its
equilibrium value at a characteristic frequency. Such dynamics are
common in structural phase transitions in which the oscillation is a
phonon of the distorted phase. Alternatively, the order parameter might
relax back to equilibrium gradually, either through dissipation or
diffusive motion of excitations. Such relaxation is influenced
by conservation laws and whether continuous symmetries are present that
support topological defects \cite{ChaikinLubensky,Goldenfeld}.
Recent time-resolved optics studies of
underdoped YBa$_2$Cu$_3$O$_{6+x}$ and La$_{2-x}$Sr$_x$CuO$_4$ showed
coherent, meV-scale, collective oscillation of the optical response
that was interpreted as an amplitude mode of the charge order
\cite{Torchinsky2013,Dakovski2015,Hinton2013}, implying
dynamics of the inertial type \cite{ChaikinLubensky,Goldenfeld}.
However, charge order is a finite-momentum phenomenon, while optics
experiments probe the system at zero momentum, so the relationship
between these oscillations and the charge order is not fully
established. Analogous experiments with momentum resolution are
therefore needed.

Here, we use time-resolved resonant soft x-ray scattering (tr-RSXS) to
study the collective dynamics of ``stripe-ordered"
La$_{2-x}$Ba$_x$CuO$_4$ with $x \sim 1/8$ (LBCO) \cite{Tranquada2004,Abbamonte2005,Hucker2011}. We use 50-fs, 1.55 eV
laser pulses to drive the charge order parameter out of equilibrium, and probe
its subsequent dynamics by scattering 60-fs x-ray pulses from a
free-electron laser (FEL) after a controlled time delay (Fig. 1A).
X-ray pulses were resonantly tuned to the Cu $L_{3/2}$ edge (933 eV)
and detected with either
an energy-integrating avalanche photodiode (APD) or an energy-resolving
soft x-ray grating spectrometer with a resolution of 0.7 eV
\cite{Doering2011,Chuang2017}. Using the latter makes this is a
time-resolved RIXS (tr-RIXS) measurement and allows isolation of the
resonant, valence band scattering from the Cu$^{2+}$ fluorescence background.
A total delay range of $T=40$ ps after the pump arrival was scanned with a time resolution of $\Delta t =$ 130 fs (SM, Fig. S1),
allowing studies of phenomena with an energy scale
ranging from $2\pi\hbar/T=0.103$ meV to
$\pi\hbar/\Delta t =15.9$ meV \cite{Abbamonte2008}.

The LBCO crystal used in this study exhibits charge order below
$T_{CO}=53$ K, which coincides with an orthorhombic-to-tetragonal
structural transition \cite{Tranquada2004,Abbamonte2005,Hucker2011}.
Experiments were carried out at $T$ = 12 K ($< T_{CO}$) and focused
around the charge order wave vector $\vec{Q}_{CO}=(0.23,0.00,1.50)$ r.l.u.,
where $(H,K,L)$ are Miller indices denoting the location of the peak in
momentum space \cite{Abbamonte2005,Hucker2011} (see SM, Materials and Methods for further details).
For a pump fluence of 0.1 mJ/cm$^2$, the energy-integrated charge order
peak, shown in Fig. 1C, is immediately suppressed due to the creation
of both electron-hole pairs and the collective excitations of interest
\cite{Forst2014,Khanna2016,Lee2012}. Fitting the momentum lineshape at
each time delay with a pseudo-Voigt function (SM, section 2), we
found that the peak is suppressed by 75\% and broadened in momentum by
45\% compared to its equilibrium profile (Figs. 1C and S3).
That the peak is not fully suppressed implies the laser provided a
perturbation of intermediate strength, where the original charge order
has not been completely extinguished. The broadening of the peak
indicates the creation of heterogeneous spatial structure in the
charge order.

It is crucial to establish whether the peak changes observed are truly
properties of the valence band. Repeating the measurement using
energy-resolved tr-RIXS with 0.7 eV resolution, we find that the peak
suppression only takes place in the resonant, quasielastic scattering
(Fig. 2). The other RIXS features, including the $dd$ and charge
transfer excitations, and Cu$^{2+}$ fluorescence emission, are
unaffected by the pump. We conclude that the effects observed are
properties of the valence band, and the time response will directly
reveal the dynamics of the charge order.

Shortly after the pump, for time $t \lesssim 2$ ps which corresponds to an
energy scale of 2 meV $\lesssim \omega \lesssim$ 15.9 meV, we observe a
shift in the wave vector of the charge order peak (see Fig. 1C). This
shift occurs in the scattering plane, along the $H$ momentum
direction, but not along the perpendicular $K$ direction (SM, Fig. S2B). A single exponential fit to the
time-dependence (Fig. 1B) indicates that the peak position recovers in
$(2.13\pm0.18)$ ps. This pump-induced phenomenon could be due to any of
three effects: (1) a change in the periodicity of the charge order, (2)
a change in the refractive index of the sample in the soft x-ray
regime, which would alter the perceived Bragg angle of the reflection
\cite{Smadici2013}, or (3) a collective recoil of the charge order
condensate.

We tested the first possibility by rotating the sample azimuthal angle
by 180$^\circ$ and repeating the measurement at the same
$\vec{Q}_{CO}=(0.23,0.00,1.50)$ r.l.u.. If the shift were due to a periodicity
change, because the CuO$_2$ plane is C$_4$-symmetric, such a rotation
would not affect the peak momentum as measured in the reference frame
of the sample. Surprisingly, we found that the momentum shift reversed
direction (Fig. 1B), meaning it is fixed with respect to the
propagation direction of the pump, not the crystal axes. This excludes
a (pure) change in the periodicity of the charge order. To test the
second possibility, we measured the $(0,0,1)$ Bragg reflection of the
LTT structure.  A pump-induced change in the refractive index should be
visible as a shift in the $(0,0,1)$ peak as well, however no such shift
was observed (Fig. S6). We are led to the surprising conclusion that
the pump induces a coherent recoil of the charge order condensate---in
essence, a nonequilibrium population of collective modes exhibiting a
nonzero center-of-mass momentum, which might be thought of as a classical
Doppler shift.

To summarize so far, the initial periodic charge order is partially destroyed
by the perturbing laser pulse, but by 2 ps, its amplitude is nearly
restored. The next stage in its approach to equilibrium is summarized in
Fig. 3A, which shows the intensity of the charge order peak for times
2 ps $\lesssim t \lesssim 10$ ps, corresponding to an energy
scale of 0.4 meV $\lesssim \omega \lesssim 2$ meV.
Unlike previous reports of an amplitude mode
\cite{Torchinsky2013,Dakovski2015,Hinton2013}, we see no coherent
oscillation indicative of inertial dynamics. We
conclude that the dynamics of the charge order in LBCO are purely
relaxational, meaning its collective modes are highly damped.

Nevertheless, the excitations of the charge order propagate
diffusively. In the standard description of relaxational dynamics
\cite{Goldenfeld,ChaikinLubensky} in a periodic system (see SM, section
7), a non-conserved order parameter driven weakly out of equilibrium
will have a time dependence proportional to $\exp(-\gamma (q) t)$,
where $q = | \vec{Q}-\vec{Q}_{CO} |$ is the momentum relative to the charge order
peak and
\begin{equation}
\gamma(q)=\gamma_0+D q^2.
\end{equation}
\noindent Here the momentum dependence arises from diffusion quantified
by the parameter, $D$ ($\gamma_0$ describes the dissipation). Fig. 3A
shows time traces of the charge order peak intensity for $t<10$ ps for a selection
of momenta $\vec{Q}$ around $\vec{Q}_{CO}$. Each curve is fit well by a single
exponential, plus a constant offset that likely arises from heating of
the electronic subsystem \cite{Perfetti2007} (see SM, section 6).
That the curves are fit well by a single exponential implies that the charge order
amplitude now deviates only weakly from its equilibrium value.
We find the relaxation rate is
highly momentum-dependent, increasing rapidly with
$q$ (Fig. 3B), and is fit well by Eq. 1, yielding dissipation parameter
$\hbar \gamma_0=(0.1730 \pm 0.0015) $ meV and diffusion
constant $\hbar D=(215 \pm 19)$ meV \AA$^2$. These two quantities imply
that the collective excitations of the charge order in LBCO propagate
by Brownian-like diffusion, with a characteristic diffusion length
$\lambda=\sqrt{2D/\gamma_0}\sim50$ \AA\ and dissipation time
$1/\gamma_0=(3.805 \pm 0.031)$ ps.

At late times, 5 ps $\lesssim t \lesssim 40$ ps, corresponding to an
energy scale of 0.1 meV $\lesssim \omega \lesssim 0.8$ meV, the order
parameter still exhibits relaxational dynamics, but its
amplitude is nearly returned to its equilibrium value. The
dynamics no longer follow a simple exponential form, but instead are
characterized by self-similar dynamic scaling
\cite{Bray2002,Mondello1990,Mondello1992,Ropers2017}. The concept of
dynamic scaling originated in the field of far-from-equilibrium
phenomena having been observed in phase ordering dynamics of
quenched binary fluids and alloys \cite{Chou1981,Gaulin1987}. The
hypothesis states that the amplitude and length scale of
the order parameter satisfy a universal relationship,
\begin{equation}
S(q,t)=L^d(t) F(q L(t)).
\end{equation}
\noindent Here, $S(q,t)$ is the time Fourier transform of $S(q,\omega)$,
$L(t)$ is the characteristic length scale of the order,
and $F(x)$ is a universal function.  For systems
with a scalar order parameter, $L$ corresponds to the mean domain size.
For systems with a continuous symmetry and a vector or tensor order
parameter, $L(t)$ corresponds to the mean distance between topological
defects, and increases as defects annihilate or are annealed from the system.
Eq. 2 states that structure on different
time scales is self-similar and independent of time if suitably scaled.
Dynamic scaling only takes place at late times following a quench when
the magnitude of the order parameter is large and nonlinear effects are
important \cite{Mondello1990,nagaya1995coarsening}.

We found that the the late-time data can be collapsed to a single curve
(Fig. 4A), using Eq. 2, taking $d=3$ and $L(t)$ as the inverse of the
half-width of the charge order reflection. This collapse implies that,
at sub-meV energy scales, the dynamics of the order parameter are
determined by universal properties such as dimensionality and ranges of
the interactions, and are not governed by any microscopic details of
LBCO itself.

The behavior of $L(t)$ is sensitive to the nature of the equilibrium
phase the system is approaching. If relaxing to a phase that is uniform
in space, $L(t)$ is known to exhibit power law behavior at long times,
$L(t) \sim t^\alpha$, where $\alpha=1/3$ if the order parameter is
conserved and $\alpha=1/2$ if it is not \cite{Mondello1990,Bray2002}.
However, if the equilibrium phase is modulated, as is the current case
of charge order in LBCO, it was predicted (for a strong quench) that
the long-time behavior of two-dimensional (2D) modulated phase order
is governed by the dynamics of topological defects, and $L(t) \sim \ln
(t)$ \cite{hou1997dynamical}.  Such slow dynamics can arise without
needing to consider additional effects due to pinning by disorder.
That neighbouring layers are correlated means the topological defects
behave as line defects oriented along the stacking direction, and so
can be described by an effective 2D dynamics.

We test these expectations by examining the long-time ($t>10$ ps)
behavior of $L(t)$.  Fig. 4B shows a compensated plot seemingly
indicating that $L(t) \sim t^{0.03}$ at long times. Small power laws of
this sort are normally interpreted as logarithmic dependence, i.e.,
$L(t) \sim \ln(t)$, supporting the prediction of Ref.
\cite{hou1997dynamical}. While disorder may be playing some role, 
this result is evidence that the long-time dynamics 
of LBCO are governed by propagation of
topological defects such as dislocation lines. 

Our study implies that the collective charge order excitations in LBCO 
are gapless down to an energy scale of 0.1 meV. The fluctuation-dissipation 
theorem implies that the charge order is fluctuating if maintained at any fixed
temperature above 1 K. On an
energy scale 0.4 meV $\lesssim \omega \lesssim 2$ meV (time scales
2 ps $\lesssim t \lesssim 10$ ps) the excitations propagate
diffusively with a mean free path $\lambda \sim 50 \AA$ in a
mean free time $1/\gamma_0=(3.805 \pm 0.031)$ ps. This implies
that an equivalent energy-domain measurement of
$S(q,\omega)$ would exhibit a featureless, quasielastic spectrum
with an energy width given by
$\gamma(q)$ in Fig. 3B \cite{Goldenfeld,ChaikinLubensky} (see SM, section 7).
At ultralow energy scales, 0.1 meV $\lesssim \omega \lesssim 0.4$ meV
(time scales 10 ps $\lesssim t \lesssim 40$ ps) the order exhibits
dynamic critical scaling, the collective excitations consisting of
topological defects whose dynamics are governed by
universal scaling laws independent of the microscopic details
of LBCO. The dynamic nature of charge order suggests it could
participate in superconductivity in a tangible way.

\section*{Acknowledgments}
We acknowledge E. Fradkin, S. A. Kivelson, T. Devereaux, B. Moritz, H. Jang, S. Lee,
J.-S. Lee, C. C. Kao, J. Turner, G. Dakovski, and Y. Y. Peng for valuable
discussions. We would also like to thank D. Swetz for helping during
the experiments and S. Zohar for his support in the data analysis. This
work was supported by U.S. Department of Energy, Office of Basic Energy
Sciences grant no. DE-FG02-06ER46285. Use of the Linac Coherent Light
Source (LCLS), SLAC National Accelerator Laboratory, is supported by
the U.S. Department of Energy, Office of Science, Office of Basic
Energy Sciences under Contract No. DE-AC02-76SF00515. M.M.
acknowledges support from the Alexander von Humboldt foundation.
P.A. acknowledges support from the Gordon and Betty Moore Foundation's EPiQS
initiative through grant GBMF-4542.

\section*{Supplementary materials}

Materials and Methods\\
Supplementary Text (sections 1-7)\\
Figs. S1 to S8\\
Table S1

\clearpage


\begin{figure}
    \centering
    \includegraphics[width=\textwidth]{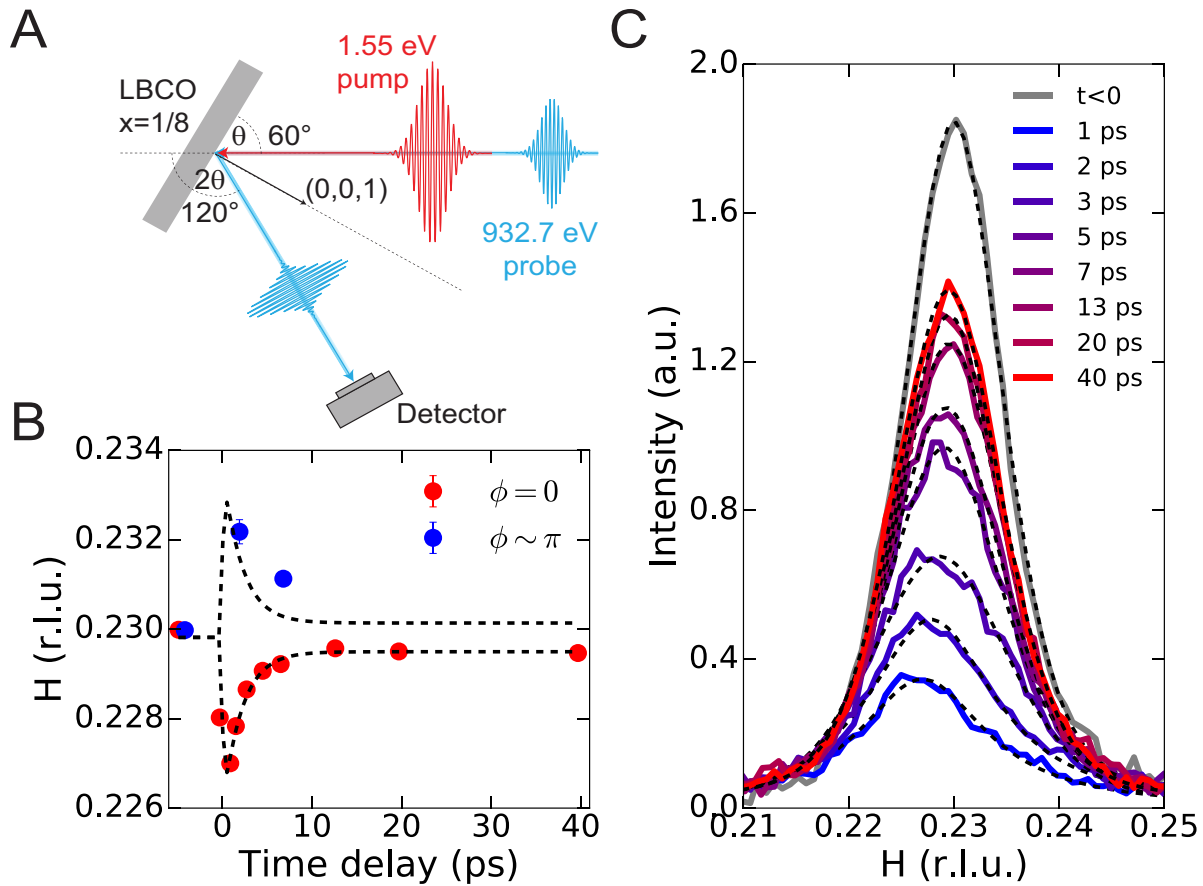}
    \caption{{\bf Pump-induced suppression and recoil of the charge order in LBCO.} (A) Layout of the experiment. 1.55 eV pump pulses perturb the charge order, which is then probed by resonant scattering of co-propagating soft x-ray FEL pulses resonantly tuned to the Cu L$_{3/2}$ edge. (B) Time-dependent shift of the charge order wavevector in the $H$ momentum direction for two different azimuthal sample angles, $\phi=0$ and $\pi$. The dashed line is a fit to the $\phi=0$ data (reflected for comparison to the $\phi=\pi$ points). (C) Momentum scan in the $H$ direction through the charge order peak for a selection of time delays. Dashed lines are fits using a pseudo-Voigt function (see SM, section 2). The fluorescence background has been subtracted.}
    \label{fig:fig1}
\end{figure}

\begin{figure}
    \centering
    \includegraphics[width=\textwidth]{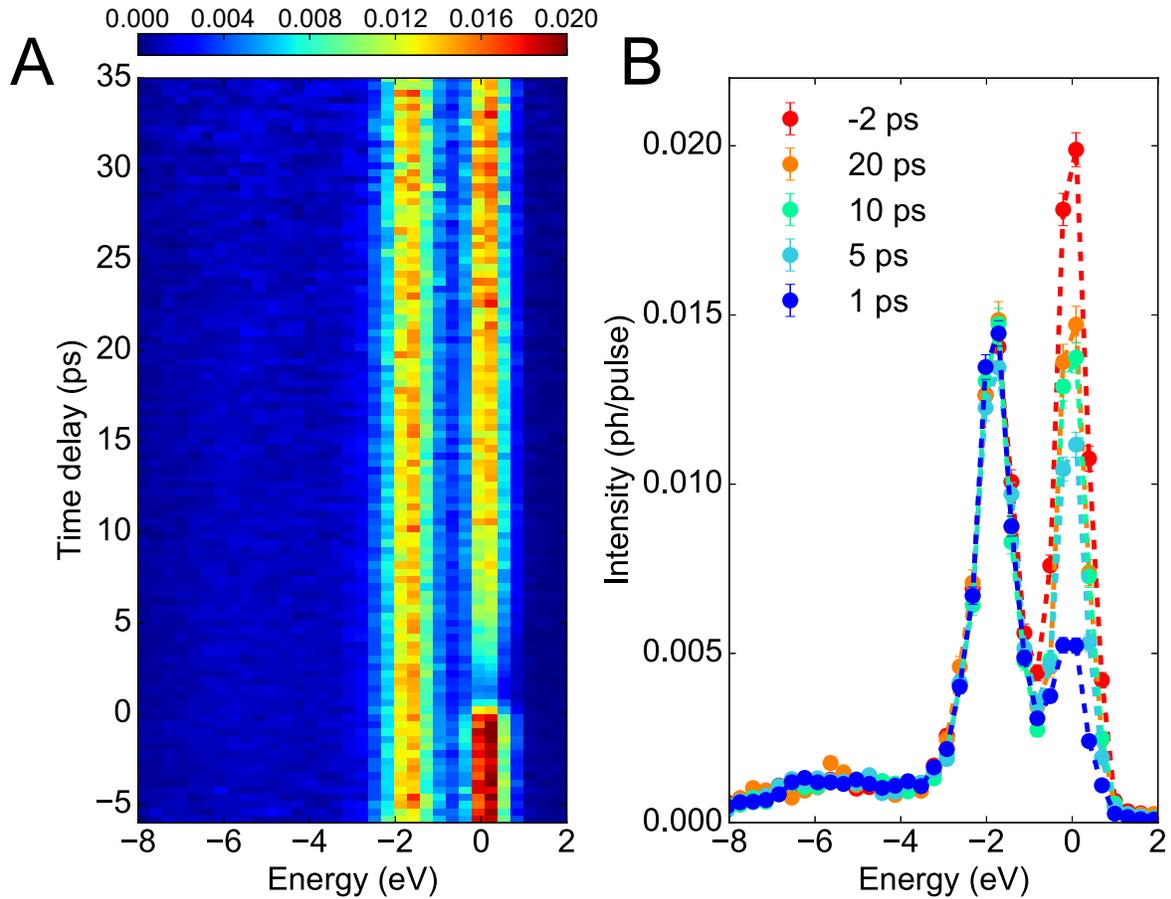}
    \caption{{\bf Time-resolved RIXS measurement of charge order in LBCO.} (A) tr-RIXS spectra taken at a
series of delay times with the momentum tuned to the peak of the charge order, Q$_{CO}$ (data are binned in 400 fs time steps to reduce counting noise in the plot). (B) Line plots of the same tr-RIXS spectra for a
selection of time delays. Error bars represent Poisson counting error. The quasielastic scattering from the charge order appears at zero energy, and is the only spectral feature influenced by the pump. The feature at -1.8 eV is a combination of $dd$ excitations and Cu$^{2+}$ emission, and the features at -6 eV are charge transfer excitations.}
    \label{fig:fig2}
\end{figure}

\begin{figure}
    \centering
    \includegraphics[width=\textwidth]{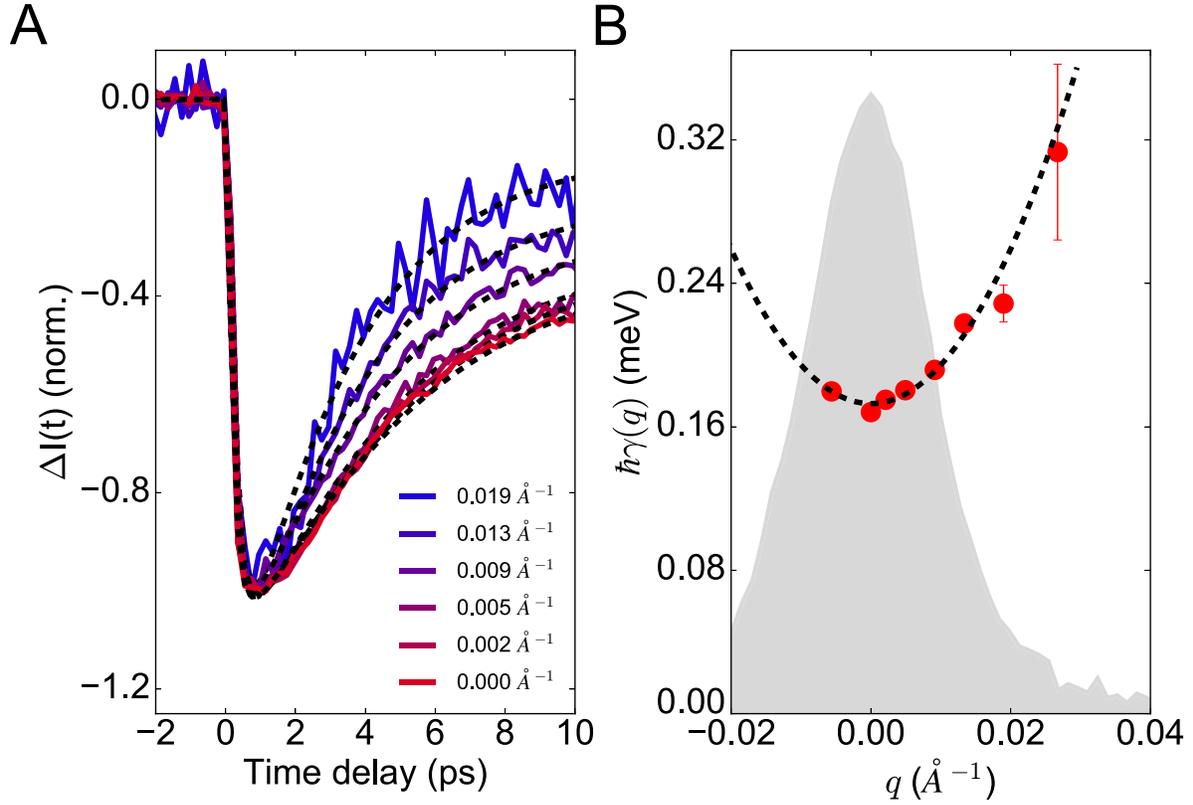}
    \caption{{\bf Collective modes of charge order in LBCO propagate diffusively.} (A) (solid lines) Time traces of the energy-integrated charge order scattering for a selection of momenta $q=|\vec{Q}-\vec{Q}_{CO}|$. The data are scaled to the same height and binned into 200 fs time steps to reduce counting noise in the plot. (dashed lines) Fits using a single exponential (see SM, section 6) show the recovery time is highly momentum-dependent. (B) (red points) Exponential decay parameter, $\gamma(q)$, as a function of relative momentum difference, $q=\textrm{sgn}(H-H_{CO})|\vec{Q}-\vec{Q}_{CO}|$. Error bars represent only the statistical uncertainties in the fits. (dashed line) Fit to the data using eq. (1). (shaded area) Lineshape of the unperturbed charge order reflection in equilibrium.}
    \label{fig:fig3}
\end{figure}

\begin{figure}
    \centering
    \includegraphics[width=\textwidth]{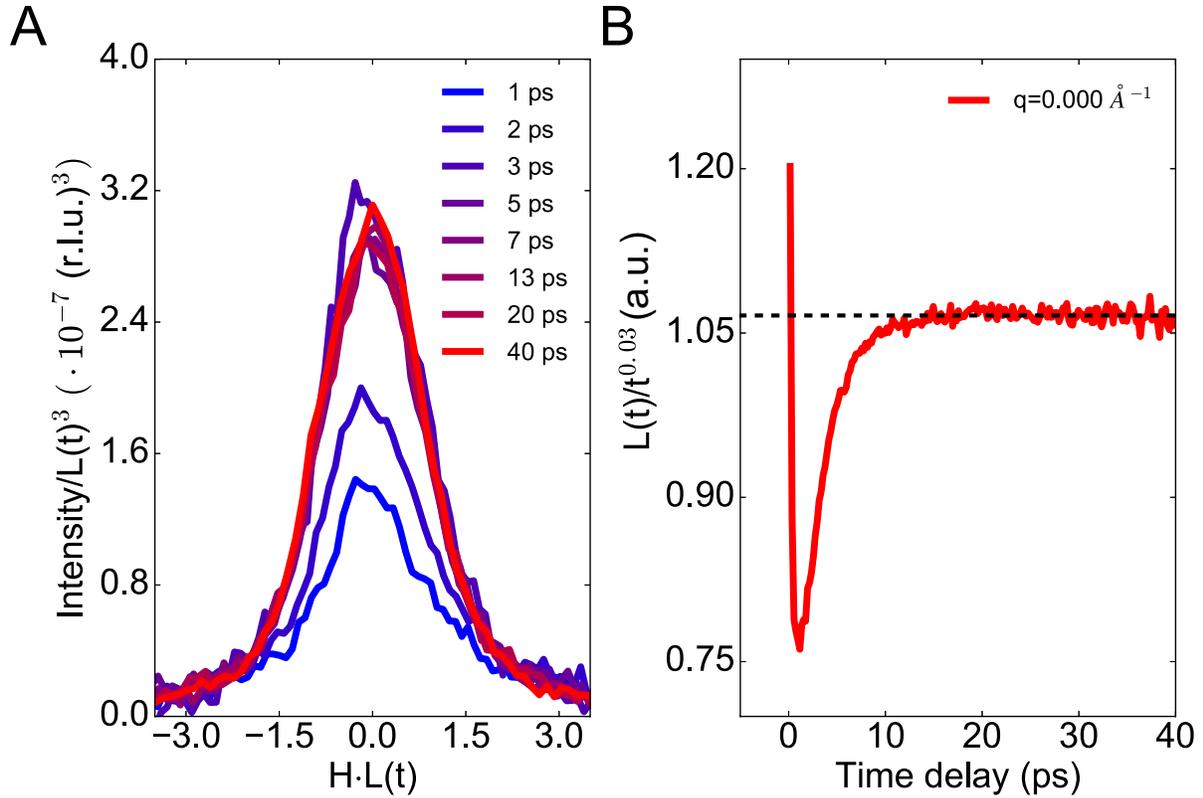}
    \caption{{\bf Demonstration of dynamic scale invariance at long times. } (A) Scaled momentum profiles (as in Fig. 1C) showing that the data collapse at late times for $d=3$. Here, $L(t)$ is taken to be the inverse half-width of the reflection at each time delay, $t$. The curves have been shifted in $H$ to compensate for the momentum recoil at short times. (B) Compensated plot of the scaling function, $L(t)$, this time taken as the cube root of the peak intensity at each time delay, $t$. The data show a power law of 0.03 at long times, indicating logarithmic behavior. }
    \label{fig:fig4}
\end{figure}

\clearpage

{\centering
\section*{\normalfont\Large{Supplementary Materials for\\Ultrafast time-resolved x-ray scattering reveals\\ diffusive charge order dynamics in La$_{2-x}$Ba$_x$CuO$_4$}}}
\vspace{2cm}
{\bf This PDF file includes:}
\begin{itemize}
\item[] Materials and Methods
\item[] Supplementary Text (sections 1-7)
\item[] Figs. S1 to S8
\item[] Table S1
\end{itemize}

\clearpage

\section*{Materials and Methods}

\subsection*{Sample growth and characterization}
A high-quality pellet of La$_{1.875}$Ba$_{0.125}$CuO$_4$ was grown by the floating zone method and cut into smaller single crystals \cite{Hucker2011}. The crystals were cleaved in air in order to expose a fresh surface, mainly oriented along the {\it ab} plane. The 2-mm-sized single crystal used in this study was pre-oriented using a lab-based Cu K$\alpha$ X-ray source. The lattice parameters were determined to be a=b=3.787 \AA\ and c=13.23 \AA\ . The surface miscut with respect to the {\it ab} crystalline plane was found to be 21 degrees. The superconducting T$_c$ of the sample was verified through a SQUID magnetometry measurement to be approximately 5 K.

\subsection*{Time-resolved Resonant Soft X-ray Scattering}
Low-temperature optical pump, soft X-ray probe measurements have been performed at the Soft X-Ray (SXR) instrument of the Linac Coherent Light Source (LCLS) X-ray free electron laser (FEL) at SLAC National Laboratory, Menlo Park, USA \cite{Schlotter2012}. The measurements reported in this work were carried out at a Resonant Soft X-ray Scattering (RSXS) endstation \cite{Doering2011} in a $3\cdot10^{-9}$ Torr vacuum. Low temperatures down to 12 K were achieved with a manipulator equipped with a Helium flow cryostat. Ultrafast probe X-rays at 120 Hz rep. rate were obtained by tuning the free electron laser to the Cu L$_{3/2}$ edge (931.5 eV) and with a 0.3 eV bandwidth after passing through a grating monochromator. The p-polarized X-ray pulses had a typical pulse duration of 60 fs, a pulse energy of 1.5 $\mu$J, and were focused down to a 1.5x0.03 mm$^2$ elliptical spot. The 1.55 eV optical pump pulses, also p-polarized, were generated with a Ti:sapphire amplifier run at 120 Hz and propagated collinearly with the X-rays into the RSXS endstation. The 50-fs pump was focused down to a 2.0x1.0 mm$^2$ spot in order to probe a homogeneously excited sample volume. The beams were spatially overlapped onto a frosted Ce:YAG crystal and synchronized by monitoring the reflectivity changes of a Si$_3$N$_4$ thin film. The shot-to-shot temporal jitter between pump and probe pulses was measured by means of a timing-tool \cite{Lemke2013,Harmand2013} and corrected by time-sorting during the data analysis. The overall time resolution of approximately 130 fs was checked by measuring the crosscorrelation signal on a polished Ce:YAG crystal (see Fig. S1). Shot-to-shot intensity fluctuations from the FEL were corrected in the photodiode data through a reference intensity readout before the monochromator. The scattered X-rays were measured with an energy-integrating avalanche photodiode located on a rotating arm at 17.3 cm from the sample, while time-resolved RIXS measurements were performed with a modular qRIXS grating spectrometer \cite{Chuang2017} mounted on a port at 135$^{\circ}$ with respect to the incident beam and provided a $\sim0.7$ eV energy resolution (FWHM) when using the 2$^{\textrm{nd}}$ order of the grating. The spectrometer was equipped with an ANDOR CCD camera operated at 120 Hz readout rate in 1D binning mode along the non-dispersive direction. The pump-probe time delay was controlled both electronically and through a mechanical translation stage. All the time-dependent rocking curves presented in this work have been referenced to their equilibrium values by selectively varying the pump-probe time delay to negative values during the data acquisition in order to minimize errors due to motor backlash and step accuracy.

\clearpage

\section*{Supplementary text}

\subsection*{1. Pump-probe cross-correlation}

\begin{figure}[h]
    \begin{center}
    \includegraphics[width=0.6\textwidth]{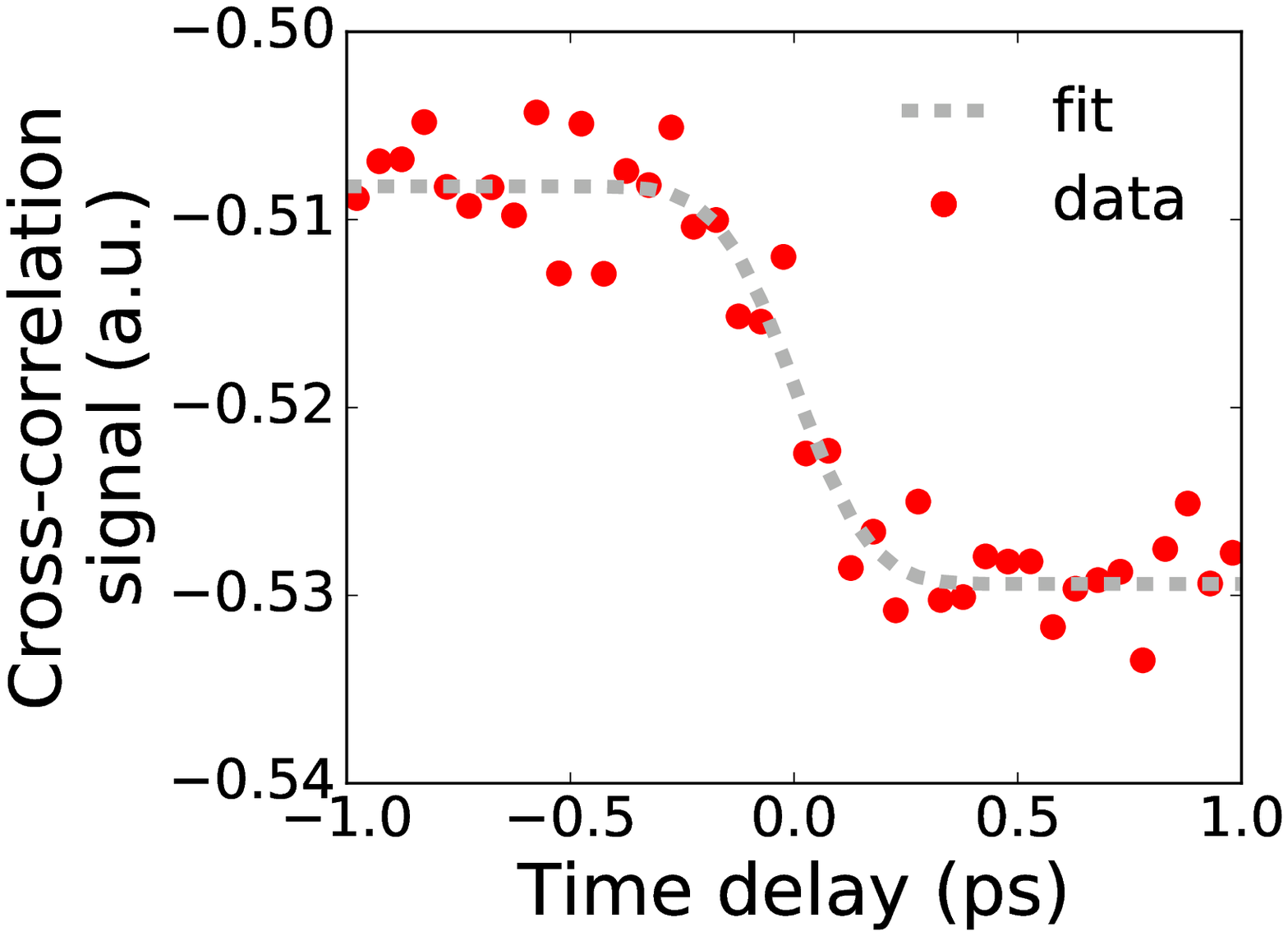}\\
    \end{center}
    {\bf Fig. S1. Optical pump-X-ray probe cross-correlation.} Time-sorted YAG transmittance edge measuring the cross-correlation between optical pump and soft X-rays at the sample position. The signal intensity is uncorrected for amplitude fluctuations of the FEL beam. The fit function is given in the text.
\end{figure}

The pump-probe cross-correlation for this experiment was measured by detecting the optical transmittance of the 1.55 eV light as a function of delay w.r.t the X-ray pulse for a 0.5 mm thick polished Ce:YAG placed at the sample position using a Si photodiode (model DET36A). When interacting with the X-ray beam, the YAG transmittance at 1.55 eV exhibits a sharp edge (see Fig. S1) that can be used to characterize the global time resolution. By fitting the signal with the function $I(t)=I_0-\frac{\Delta I}{2}\big[1+\textrm{erf}\big(\frac{t}{\tau_0}\big)\big]$, we obtain a cross-correlation width $\tau_0=(187\pm39)$ fs. Hence, our global time-resolution is $\tau_0/\sqrt{2}\sim130$ fs under the assumption of Gaussian beam envelopes.

\subsection*{2. Charge order peak rocking curves fit and background subtraction}

The rocking curves shown in Fig. S2 have been fitted with a pseudo-Voigt profile and a linear background,
\begin{equation}
I(q)\bigg|_t=(I_0+mq)+f\frac{1}{\pi g}\frac{A}{1+\bigg(\frac{q-Q_{CO}}{g}\bigg)^2}+(1-f)\frac{A}{g}\sqrt{\frac{\ln2}{\pi}}\exp\bigg[-\ln2\bigg(\frac{q-Q_{CO}}{g}\bigg)^2\bigg]
\end{equation}
The first term represents the linear background, while the second and third term represent a Lorentzian and a Gaussian, respectively, with $f$ as a linear mixing parameter. The last two terms share the same amplitude parameter $A$ and the same FWHM $2g$.
The fit parameters at each time delay along the H direction (Fig. S2) are reported in Fig. S3. The background slope and intercept are constant over the entire delay window, therefore a background subtraction based on the fitted slope does not introduce artifacts in the time-dependent behavior of the charge-order (CO) peak. The fit parameters in Fig. S3A-D exhibit a time dependence that can be captured by a single exponential recovery and an offset, while the background is time-independent.
The same fit procedure has been also applied for the scans along the K direction shown in Fig. S2B. The fit parameters for those curves are reported in Tab. S1.

\begin{center}
\begin{tabular}[h!]{ c c c }
 \hline
 & t$<$0.0 ps & t=1.0 ps\\\hline
A (a.u.) & $(14.02\pm0.34)\cdot10^{-3}$ & $(6.35\pm0.28)\cdot10^{-3}$ \\
 $K_{CO}$ & $(0.0\pm0.3)\cdot10^{-4}$ & $(-7.4\pm0.5)\cdot10^{-4}$ \\
 g (r.l.u.) & $(4.03\pm0.05)\cdot10^{-3}$ & $(5.23\pm0.09)\cdot10^{-3}$ \\
 f & $(0.655\pm0.048)$  & $(0.652\pm0.080)$ \\
 m (r.l.u.) & $(0.00\pm0.18)$ &$(0.00\pm0.10)$ \\
 $I_0$ (a.u.) & $(2.3031\pm0.0057)$ &$(2.3168\pm0.0042)$\\
 \hline
\end{tabular}\\
\vspace{12pt}
\baselineskip16pt
{\bf Table S1. Fit parameters along K projection.} Pseudo-Voigt fit parameters for the curves in Fig. S2B. $K_{CO}$ is the Miller index of the $Q_{CO}$ wavevector in the fit expression.
\end{center}

\begin{figure}[h!]
    \begin{center}
    \includegraphics[width=0.7\textwidth]{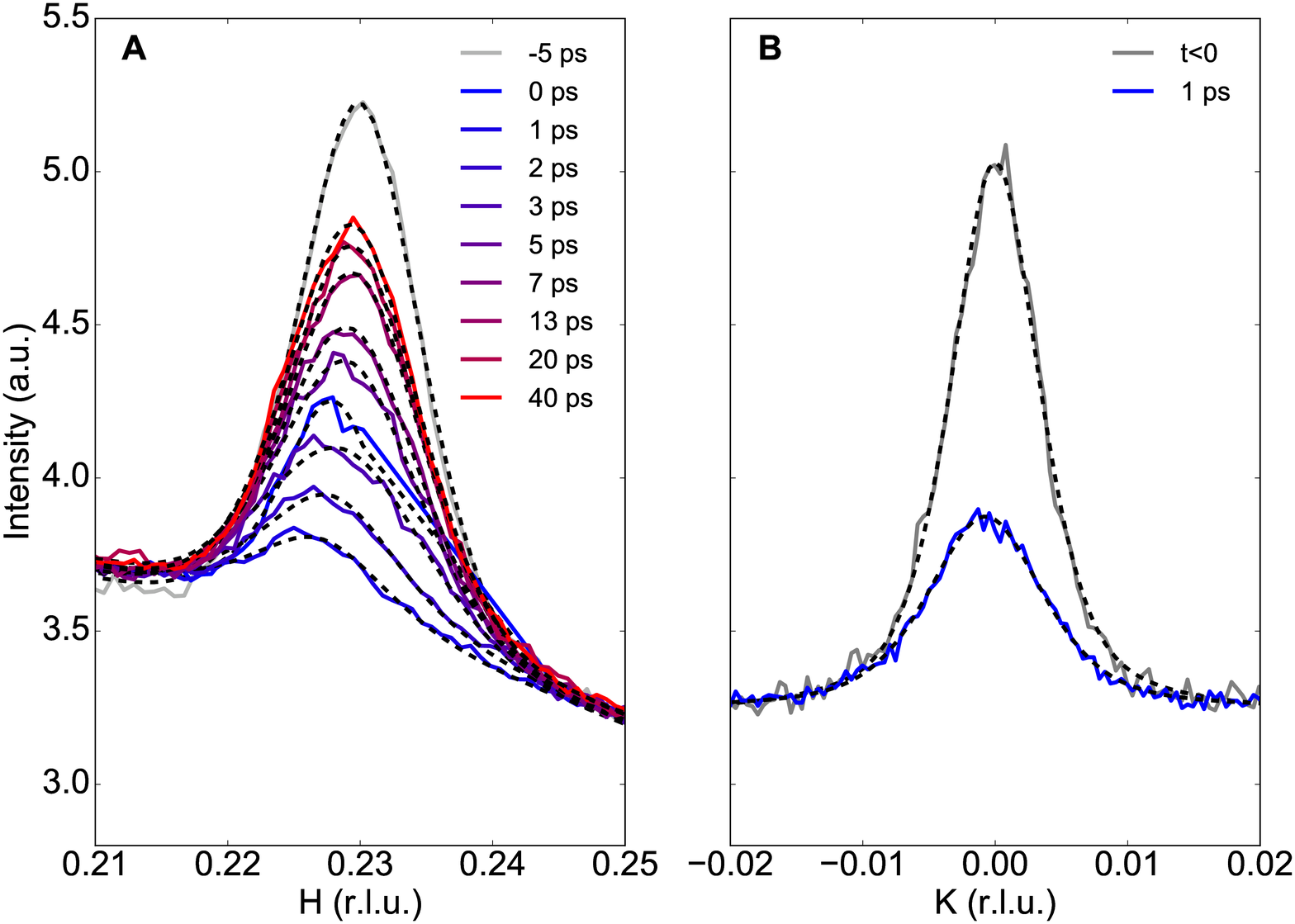}\\
    \end{center}
    {\bf Fig. S2. Pump-induced CO peak melting.} CO peak projection along H and K for selected time delays. Solid lines are experimental data, dashed lined represent pseudo-Voigt fits.
\end{figure}

\begin{figure}[h!]
    \begin{center}
    \includegraphics[width=0.9\textwidth]{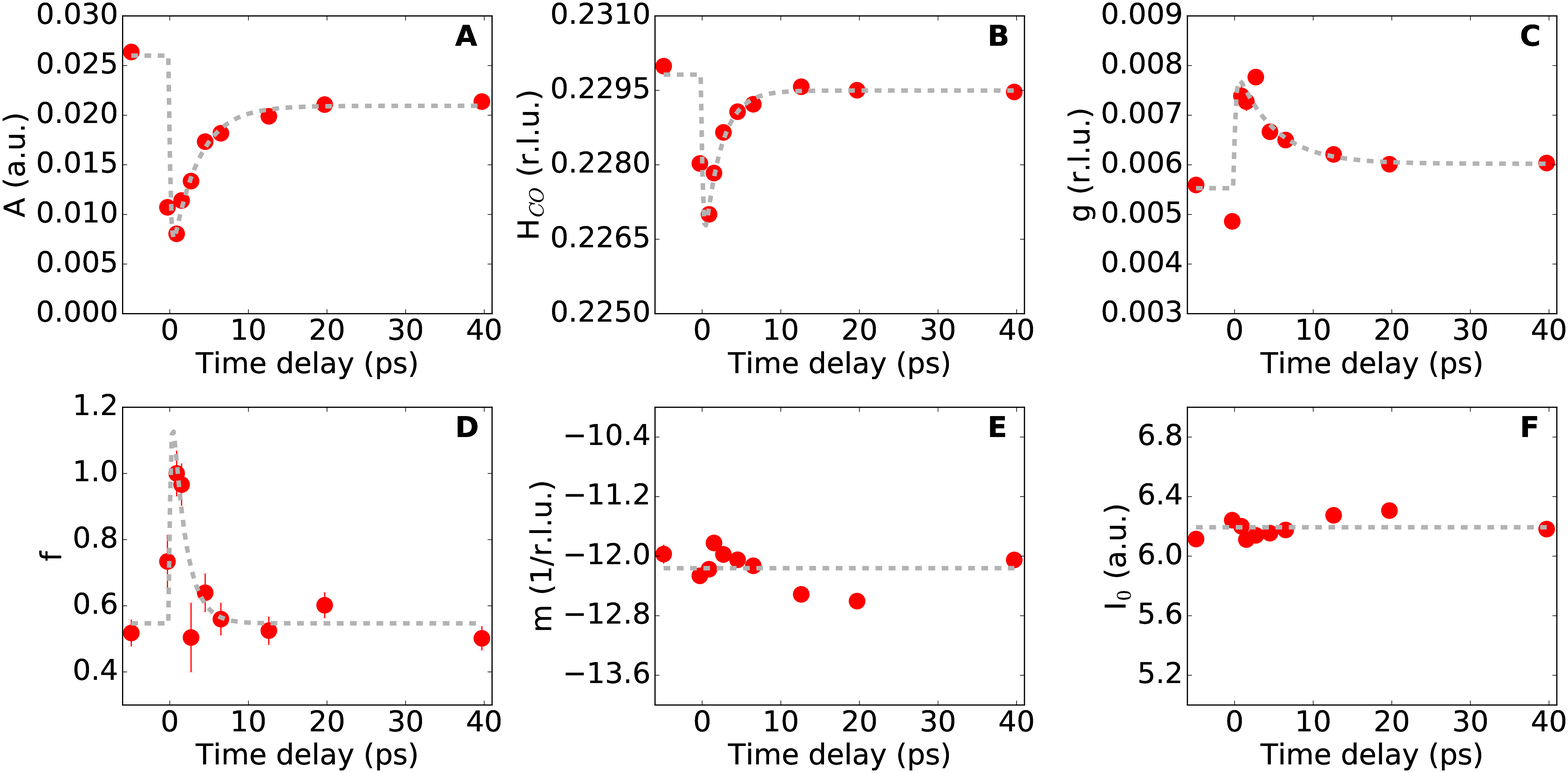}\\
    \end{center}
    {\bf Fig. S3. Time-dependent CO peak fit parameters.}  (A) Pseudo-Voigt amplitude A, (B) H projection of the CO wavevector Q, (C) CO peak HWHM g, (D) Pseudo-Voigt mixing parameter f, (E) linear background slope m (F) linear background intercept I$_0$. Red symbols mark the fit parameters. These points are then fit to an exponential decay in (A)-(D) (dashed grey lines) and to a constant in (E)-(F).
\end{figure}

\clearpage

\subsection*{3. Charge order peak rocking curves for $\phi\sim\pi$}

\begin{figure}[h!]
    \begin{center}
    \includegraphics[width=0.7\textwidth]{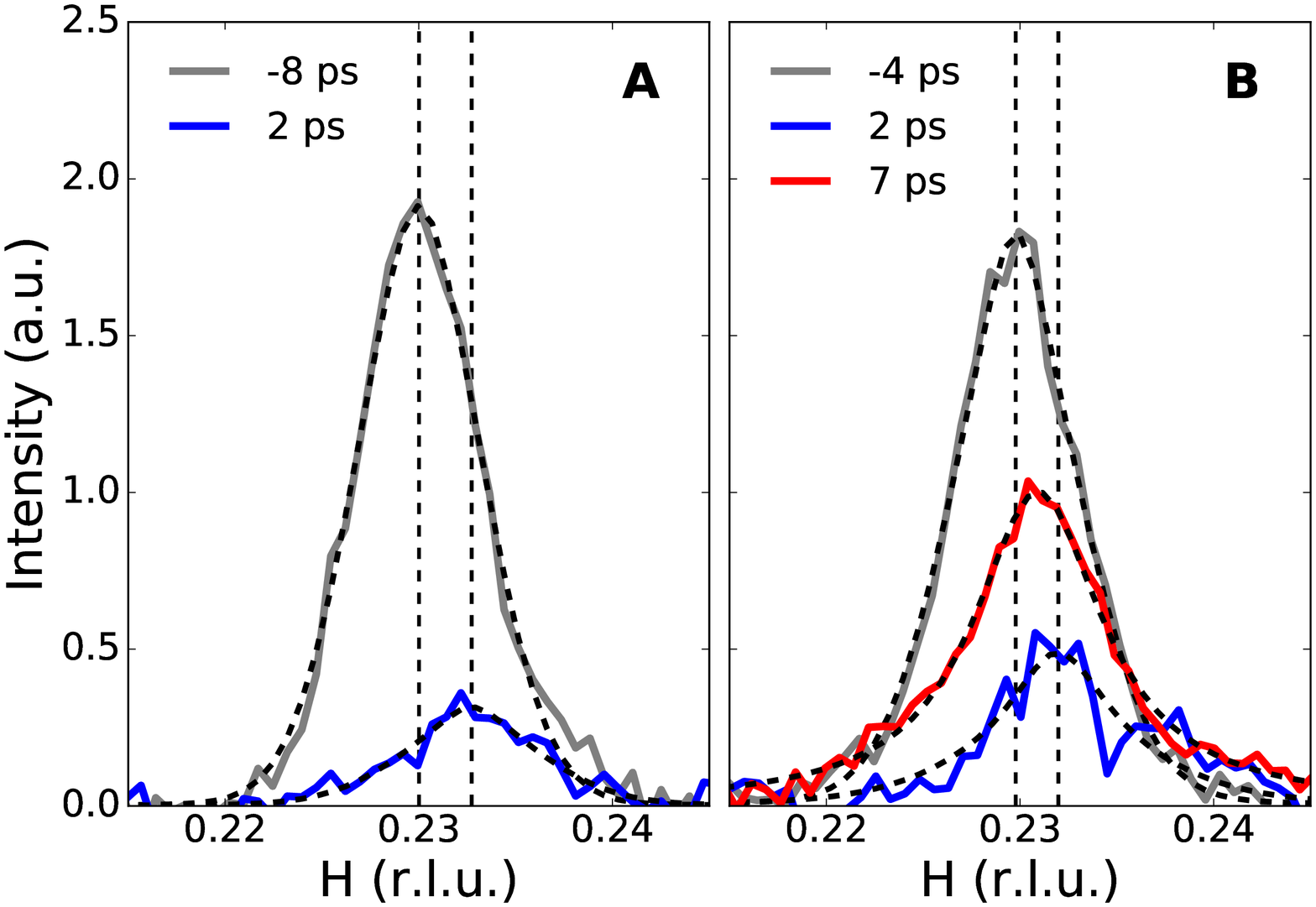}\\
    \end{center}
    {\bf Fig. S4. CO peak shift at $\phi\sim\pi$.} CO peak projection along H for selected time delays. Solid lines are experimental data, dashed lines represent pseudo-Voigt fits. Vertical dashed lines indicate the peak positions at equilibrium and at the maximum of the response. Data shown in (A) are acquired with a pump fluence of 0.2 mJ/cm$^2$ while data in (B) with 0.1 mJ/cm$^2$.
\end{figure}

In the main text, we discuss the change in the CO peak shift direction when rotating the sample around the azimuthal angle $\phi$. In Fig. S4, we show the CO rocking curves (with background subtraction) for the blue points in Fig. 1B of the main text. These data are acquired at the same pump fluence and temperature conditions as the data reported in the rest of Fig. 1. Moreover, here we also report a second dataset exhibiting a clear peak shift when irradiated with a higher IR pump fluence.

\clearpage

\subsection*{4. Comparison between tr-RIXS and APD data}

The time-dependent elastic line intensity measured with the RIXS spectrometer and integrated over the energy axis maps closely onto the time-dependent, background-subtracted CO peak intensity measurement carried out with the avalanche photodiode (see Fig. S5).

\begin{figure}[h!]
    \begin{center}
    \includegraphics[width=0.6\textwidth]{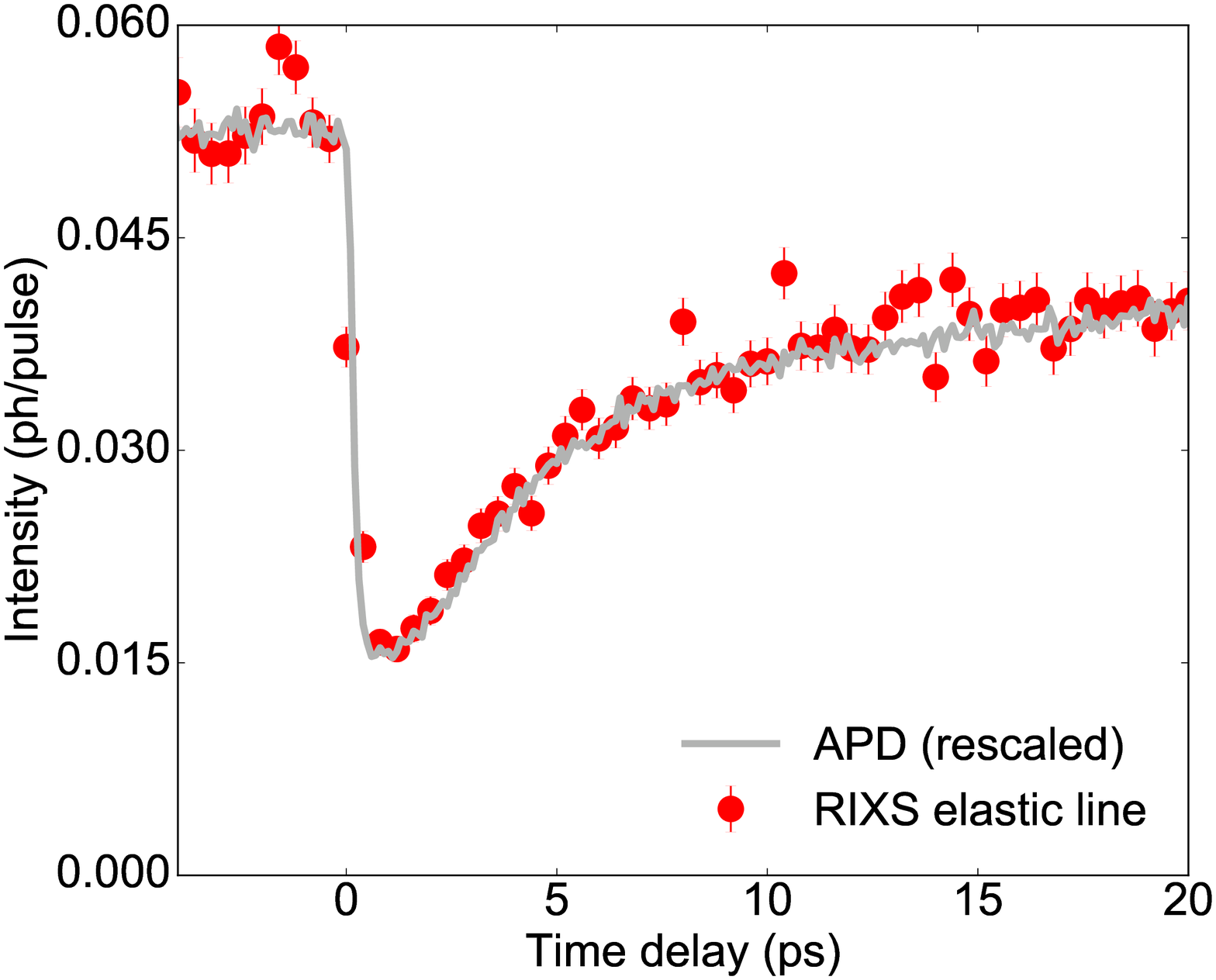}\\
    \end{center}
    {\bf Fig. S5. Comparison between tr-RIXS and energy-integrated time dependence at Q$_{CO}$.}  Elastic line intensity values (energy-integrated in a 1.5-eV range around the peak) vs time delay are reported as red dots, while a rescaled, background-subtracted photodiode (APD) intensity measurement at Q$_{CO}$ is shown as a solid grey line. Error bars are Poisson uncertainties.
\end{figure}

\subsection*{5. Response of the LTT distortion peak}

The onset of CO in 1/8-doped LBCO is accompanied by a low-temperature structural transition from a low-temperature orthorhombic (LTO) to a low-temperature tetragonal (LTT) phase \cite{Hucker2011}. This structural change allows us to observe an otherwise forbidden (0,0,1) Bragg reflection. The (0,0,1) reflection measurement is performed at resonant condition with Cu L$_3$ edge X-rays. Previous studies reported pump-induced changes in this peak under 1.55 eV \cite{Khanna2016} and midinfrared \cite{Forst2014} irradiation and for mJ/cm$^2$ fluences. Hence it is important to check whether the structure responds as well to the excitation in the current experimental conditions. The (0,0,1) peak data at the same fluence of the CO data shown in the main text are shown in Fig. S6. The peak intensity decreases but the peak does not shift in Q, at variance with the CO diffraction signal. This is additional evidence that the lattice does not change while the charge moves.

\begin{figure}[h]
    \begin{center}
    \includegraphics[width=0.4\textwidth]{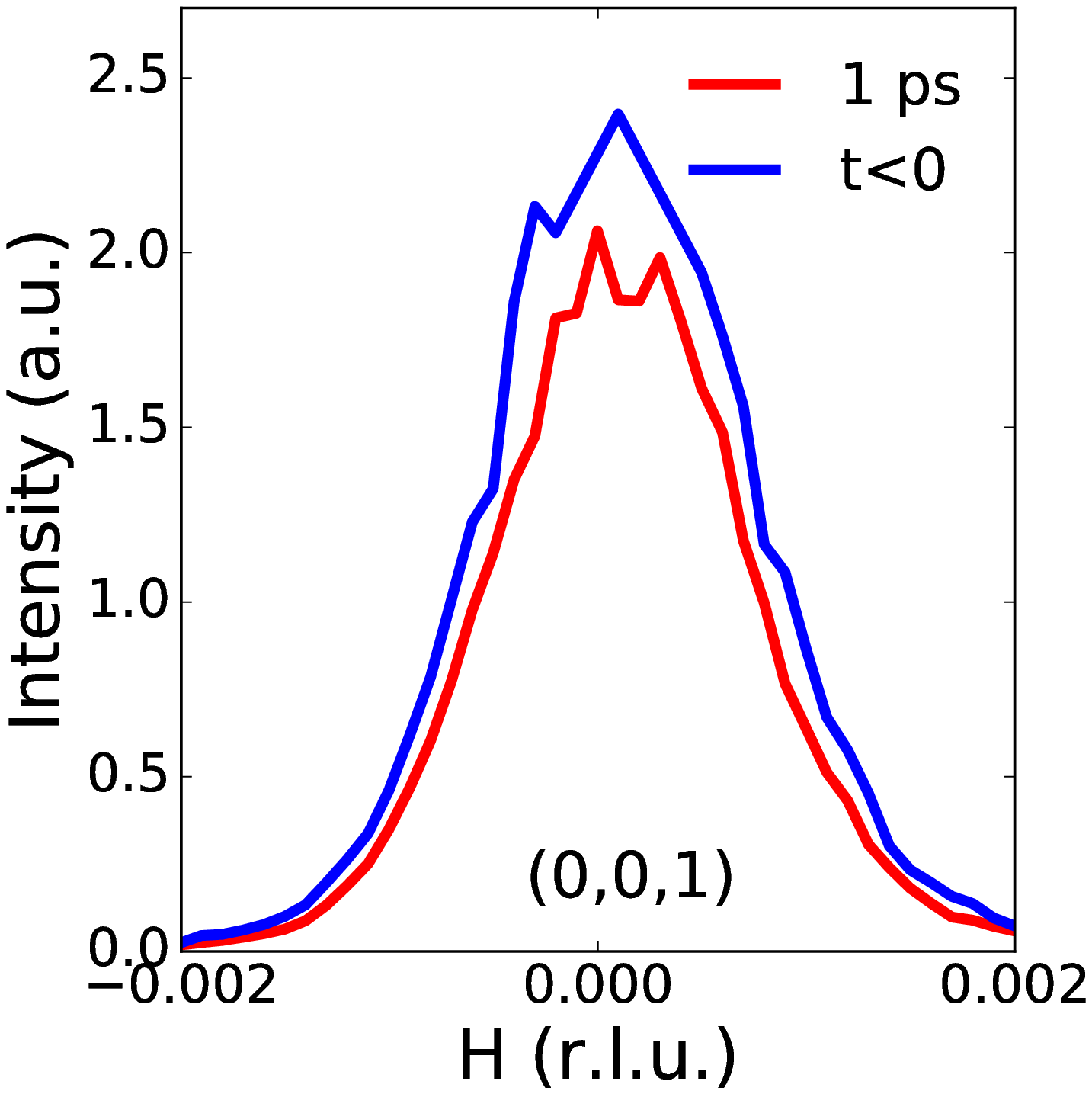}\\
    \end{center}
    {\bf Fig. S6. Dynamics of the LTT distortion.} Projection of the (0,0,1) Bragg peak along the H direction at T=12 K for selected time delays and for a pump fluence of 0.1 mJ/cm$^2$.
\end{figure}

\subsection*{6. Raw time-dependent, energy-integrated peak intensities around $Q_{CO}$}
In Fig. 3A of the main text, we show normalized differential intensity changes vs pump-probe time delay. In Fig. S7 we show the unscaled intensity curves after time-sorting and rebinning with 200 fs time steps. Shot-to-shot intensity fluctuations have been corrected with a reference intensity monitor prior to the monochromator, while the fluorescence background has not been subtracted out. Each intensity curve is fit with a single exponential recovery and an offset capturing the long-time relaxation of the CO peak. The fit function for each momentum cut is
\begin{equation}
I(t)\Bigg|_q= \bigg|\Delta_0-\Theta(t-t_d)\bigg[1-\exp\bigg(-\frac{t-t_d}{\tau_0}\bigg)\bigg]\bigg[A_0+A_1\exp\bigg(-\frac{t-t_d}{\tau}\bigg)\bigg]\bigg|^2+I_{\textrm{bkg}}.
\end{equation}
In this expression, the pump-induced signal grows with an exponential saturation characterized by a timescale $\tau_0$. $A_1$ and $A_0$ respectively represent the exponential amplitude and an offset, while $\tau$ is the timescale of the exponential recovery. $t_d$ is the parameter for the zero time delay. $\Delta_0$ is the equilibrium value of the order parameter, while $I_{\textrm{bkg}}$ is the fluorescence contribution to the overall intensity. The fit curves are indicated in Fig. S7 as black dashed lines.
\begin{figure}[h]
    \begin{center}
    \includegraphics[width=0.6\textwidth]{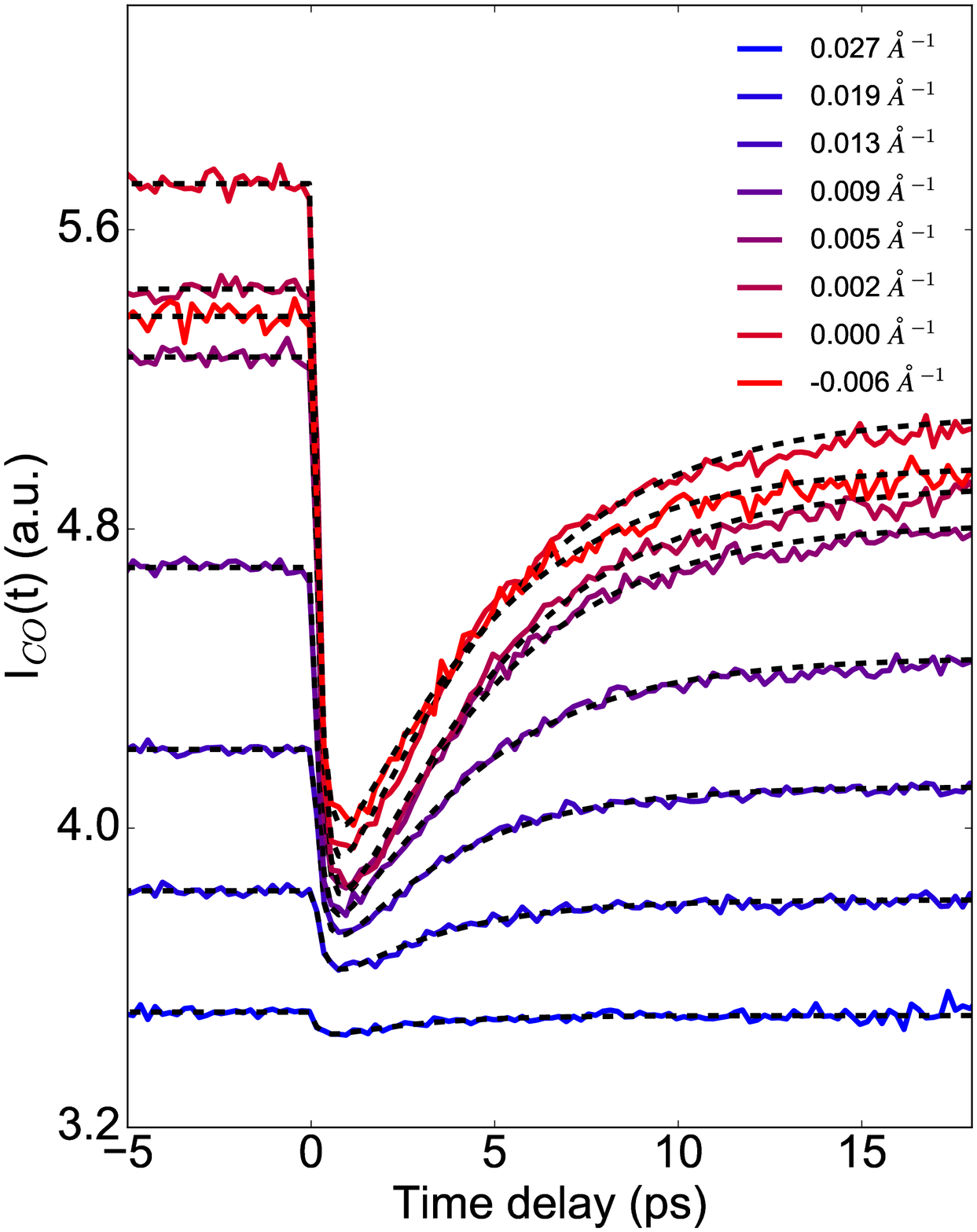}\\
    \end{center}
    {\bf Fig. S7. Raw time-dependent CO peak intensity.} Energy-integrated, time-dependent intensity profiles of the CO peak for a pump fluence of 0.1 mJ/cm$^2$ and for selected momenta (solid lines). Data are binned along the time axis in 200 fs steps to improve statistics. Single exponential fit curves are shown as dashed black lines.
    \label{fig:figS6}
\end{figure}

\clearpage

\subsection*{7. Momentum-dependent recovery of the charge-ordered phase}

The purpose of this section is to calculate the relaxation of
long-wavelength fluctuations of the charge order parameter as it
approaches an equilibrium periodic state, in the spirit of Landau
theory.  The calculation remains in the framework of the time-dependent
Ginzburg-Landau equation but the coarse-grained free energy takes into
account the periodic charge order state.

Our data indicate that following the application of a laser pulse, the
periodic state undergoes exponential relaxation with a rate $\gamma
(\textbf{q})$ in the interval 2 ps $ < t < 10$ ps, where $\textbf{q}$ is
the momentum relative to the charge order peak at $\textbf Q_{CO}$. Our
goal is to calculate the functional form of $\gamma (\textbf{q})$, and
show that it is of the form of Eq. (1) in the main text.

We model the charge density condensate as a function of space
$\textbf{x}$ and time $t$ by the Swift-Hohenberg equation, a widely
used minimal model of periodic pattern formation
\cite{swift1977hydrodynamic,cross1993pattern}, which has previously
been used to describe charge density wave dynamics
\cite{karttunen1999defects}.  Written in canonical form it is:
\begin{equation}
\tau_0\frac{\partial\psi}{\partial t} = \epsilon \psi - {\psi}^3 - \xi_0^4(q_0^2 + \nabla ^2)^2 \psi
\label{SHeqn}
\end{equation}
Here $\psi = \psi(\textbf{x}, t)$ is the charge order parameter
rescaled so that the cubic term has coefficient unity, $\tau_0$ is a
relaxation time, $q_0$ is the magnitude of the wavevector at the onset
of ordering when the control parameter $\epsilon > 0$ and $\xi_0$ is a
charge fluctuation correlation length. The control parameter is
determined by the degree to which the temperature is below the critical
temperature for charge ordering.  In the experiment, apart from the
period when the system is excited by the laser, the temperature is well
below the critical temperature, and we are deep into the regime where
periodic charge order occurs.  In principle this equation should have
an additive noise that obeys the fluctuation-dissipation theorem, but
this will not concern us if we restrict our calculation purely to
linear stability. Note that the Swift-Hohenberg equation does not
conserve the charge, and this is appropriate because it is only the
total charge from the condensate and the quasi-particles that is
conserved.

When $ \epsilon > 0$ the uniform state $\psi=0$ becomes linearly
unstable to the formation of a periodic state.  The wavelength of this
periodic state is not unique, because there is a band of linearly
stable periodic steady states around the most unstable mode with
wavenumber $q_0$, thus posing the so-called pattern selection problem.
A large body of work shows in detail how the initial and boundary
conditions as well as the history of the system determine which of
these possible steady states is actually chosen by the dynamics
\cite{cross1993pattern}, but here we simply use the observed charge
order state rather than try to predict what it should be.  Given that a
periodic state of charge order exists, the next step is to perform the
linear stability analysis around this pattern.

We assume the stripe patterns are periodic in the $x$ direction, and
constant in the $y$ direction, and make a single-mode approximation:
\begin{equation}
\psi(\textbf{x},t) = \psi_0 \left[A(x,y,t)e^{iq_0 x} + c.c.\right]
\end{equation}
where the slowly-varying complex amplitude $A$ can be shown to obey the equation
\cite{cross1993pattern}
\begin{equation}
\partial_t A = \epsilon A + \mu_0^2 \left(\partial_x - \frac{i}{2q_0}\partial_{y}^{2} \right)^2 A - 3|A|^2 A
\end{equation}
where $\mu_0 = 2q_0$.  It is known that the single-mode approximation
is qualitatively accurate and that the dependence with $\epsilon$ of
the selected wave vector is weak, so we use this amplitude equation
description as a first approximation to describe the long-wavelength
dynamics.

Rescaling the equation by $q_0 \rightarrow \frac{q_0}{\mu_0}$, $A \rightarrow A/\sqrt{3}$ and $\textbf{x}
\rightarrow \mu_0 \textbf{x} $, we obtain the Newell-Whitehead equation
in canonical form \cite{newell1969finite}:
\begin{equation}\label{eq:amp}
\partial_t A = \epsilon A + \left(\partial_x - \frac{i}{2q_0}\partial_{y}^{2} \right)^2 A - |A|^2 A
\end{equation}

Since the wavenumber of the stripes can be anywhere within the band,
and is determined through an pattern selection process that is not of
concern here, we will denote the wavenumber of the actual selected
stripe pattern to be $q_0 + k$.  This means that our stability analysis
is around the state described by the complex amplitude
\begin{equation}
A_k = a_k e^{ikx},
\end{equation}
where $a_k=\sqrt{\epsilon-k^2}$.
We now impose a small perturbation, i.e. $A=A_k + \delta A$ with wavenumber $\textbf{q} = (q_x, q_y)$ where:
\begin{equation}
\delta A = (\delta a_{+} e^{i \textbf{q}\cdot\textbf{x}} + \delta a_{-} e^{-i \textbf{q}\cdot\textbf{x}}) e^{ikx}
\end{equation}
and the negative mode is included due to the presence in
the linearized equation of the term $\delta A^*$.

Plugging back into the Newell-Whitehead equation, we find
the linearized equations of motion:
\begin{equation}\label{eq:coupled}
\partial_t \delta a_+ = -(a_k^2 + U_+)\delta a_+ - a_k^2 \delta a_- \\
\partial_t \delta a_- = - a_k^2 \delta a_+ - (a_k^2 + U_-)\delta a_-
\end{equation}
where
\begin{equation}
U_\pm = (k\pm q_x)^2 + \frac{(\pm q_x + k)q^2_y}{q_0} - \frac{q^4_y}{4q_0^2} - k^2
\end{equation}
By writing eq. (\ref{eq:coupled}) in matrix form and solving for the eigenvalues $\lambda_{\pm}$, we obtain:
\begin{equation}
\begin{array}{cl}
\lambda_\pm
&= - a_k^2 - \frac{U_+ + U_-}{2} \pm\sqrt{a_k^4 + \frac{(U_+ - U_-)^2}{4}} \\
&= - \epsilon + k^2 - q_x^2 -  \frac{k q_y^2}{q_0} + \frac{q_y^4}{4q_0^2} \pm \sqrt{(\epsilon-k^2)^2 + 4k^2 q_x^2 + \frac{4kq_x^2 q_y^2}{q_0} + \frac{q_x^2 q_y^4}{q_0^2}}\\
&\simeq - \epsilon + k^2 - q_x^2 \pm \sqrt{(\epsilon-k^2)^2 + 4k^2 q_x^2} \\
&\simeq - \epsilon + k^2 - q_x^2 \pm \left[(\epsilon-k^2) + \frac{2k^2 q_x^2}{\epsilon-k^2}\right]
\end{array}
\end{equation}

In the scattering geometry of our experiment, sketched in Fig. S8, $q_y \ll q_x$ so that we
find to a good approximation
\begin{equation}\label{eigenvalues}
\lambda_+ =  - \frac{\epsilon - 3k^2 }{\epsilon-k^2} q_x^2, \qquad
\lambda_- =  -2(\epsilon-k^2) - \frac{\epsilon + k^2}{\epsilon-k^2} q_x^2
\end{equation}

\begin{figure}[h!]
    \begin{center}
    \includegraphics[width=0.6\textwidth]{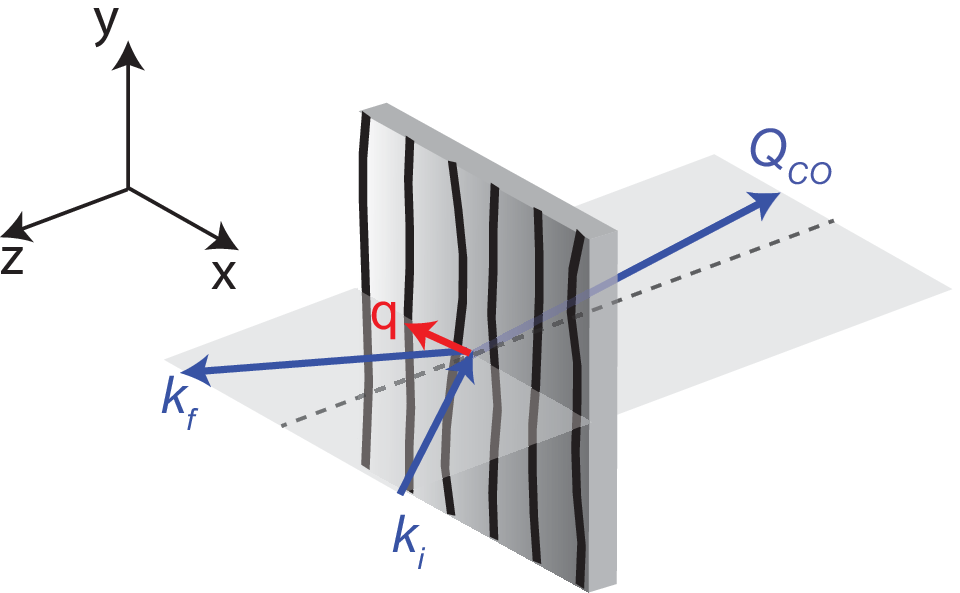}\\
    \end{center}
    {\bf Fig. S8. Schematic representation of the scattering geometry.}
    For the purpose of this section, the xy plane is defined as
    parallel to the CuO$_2$ planes. The scattering plane lies
    orthogonal to the stripe direction, here denoted as y. $k_i$ and
    $k_f$ represent the incident and scattered momenta of the x-ray
    beam, $Q_{CO}$ is the charge order wavevector described in the main
    text and $q$ is the small momentum deviation from $Q_{CO}$
    considered in this section.
    \label{fig:figS8}
\end{figure}

The mode with eigenvalue $\lambda_+$ is a Goldstone phase mode that in
the limit of vanishing $q_x$ restores translational invariance.  It is
likely not to be present in our system because of disorder or grain
boundaries between the stripe domains, both of which break
translational invariance.

The mode with eigenvalue $\lambda_-$ is a decaying mode that
corresponds to that measured in Fig. 3B of the main text.  We conclude by rewriting it in
the physical units.  The $x$-component of the measured charge order
wavevector is $Q_{CO} = q_0 +k$ and $k \rightarrow k \cdot
2q_0\cdot\xi_0^2$.  Since $\textbf{q}$ is defined relative to the
charge order wavevector, we have that $q=q_x + O(q_y/q_x) \approx q_x$.
Further neglecting the dependence of $k$ on $\epsilon$ leads to the simple
formulae for the decay rate extracted from the experiment:
\begin{equation}
\gamma (q) = 2\epsilon+ q^2
\end{equation}
or in the original units:
\begin{equation}
\gamma (q) =  \frac{2\epsilon}{\tau_0}+ \frac{(2q_0 \xi_0^2)^2}{\tau_0} q^2.
\end{equation}
This result is Eq. (1) of the main text.

\clearpage

\bibliography{LBCO-DynamicScaling-v8}

\bibliographystyle{Science}

\end{document}